\documentclass[preprint,review,times,3p] {elsarticle}
\usepackage{graphicx}
\usepackage{subfig}
\usepackage[countmax]{subfloat}
\usepackage{amsmath, amssymb}

\begin{document}

\begin{frontmatter}

\title{Theoretical Investigation of Local Electron Temperature in Quantum Hall Systems}

\author[label1]{N. Boz Yurda\c{s}an\corref{label4}}
\author[label1]{K. Akg\"ung\"or}
\author[label2,label3]{A. Siddiki}
\author[label1]{\.I. S\"okmen}

\address[label1] {Dokuz Eyl\"{u}l University, Physics Department, Faculty of
  Arts and Sciences, 35160 Izmir, Turkey }
\address[label2] {Istanbul University, Physics Department, Beyazit Campus, \.Istanbul,
Turkey}
\address[label3]{Department of Physics, Harvard University, Cambridge, MA 02138, USA}

\begin{abstract}
In this work we solve thermo-hydrodynamical equations considering a two dimensional electron system in the integer quantum Hall regime, to calculate the spatial distribution of the local electron temperature. We start from the self-consistently calculated electrostatic and electrochemical potentials in equilibrium. Next, by imposing an external current, we investigate the variations of the electron temperature in the linear-response regime. Here a local relation between the electron density and conductivity tensor elements is assumed. Following the Ohm's law we obtain local current densities and by implementing the results of the thermo-hydrodynamical theory, calculate the local electron temperature. We observe that the local electron temperature strongly depends on the formation of compressible and incompressible strips.

\end{abstract}

\end{frontmatter}

\section{Introduction}

\label{intro} Quantum Hall effects (QHE's) are remarkable phenomena observed at two-dimensional electron systems (2DESs), where the longitudinal resistance $R_{xx}$ vanishes in the low current regime while the Hall resistance $R_{xy}$ is quantized to one over integer or fractional multiples of $h/e^{2}$, where $e$ is the elementary charge and $h$ is the Planck's constant \cite{Klitzing80:449,FQHE}. If the current (density) is increased up to a critical value, $R_{xx}$ increases by several orders of magnitude within a narrow range of the current amplitude and this observation is called the breakdown of the QHE \cite{Ebert83:5441,Cage83:1374,Kuchar84:196,Kaya98:7536,Akera2002:228}. To investigate the microscopic nature of the breakdown phenomena one essentially has to know the current distribution, i.e. where the Hall potential drops across the Hall bar. This information is provided by local probe experiments, where the spatial distribution of the induced (Hall) potential is measured by scanning force microscope techniques~\cite{Weitz00:247,Dahlem:2010:121305}. It is observed that, the spatial distribution of the Hall potential strongly depends on the applied magnetic field $B$, such that if the system is out of the plateau regime Hall potential varies linearly across the sample resembling the classical Hall effect. In contrast, within the plateaus, the potential presents strong non-linear distributions: At the high $B$ field side of the plateau the potential drop is spread all across the sample indicating that the current is carried at the bulk, meanwhile at the low field side one observes sharp variations at the opposing edges, whereas the potential is constant at the bulk. This behavior is attributed to edge state transport. The spatial variation of the potential drop as a function  of $B$ field shows that, while decreasing the field the edge states move towards the physical ends of the sample until the plateau disappears. A simple calculation of the spatial distribution of the edge states, following the original work of Chklovskii et al, shows that the position of the potential drop almost perfectly matches with the position of the edge states. The theoretical work takes into account electron-electron interactions within a Thomas-Fermi approximation and provides estimations of the widths and the positions of the compressible and the incompressible strips, for a fixed depletion length. There, it is assumed that the 2DES is separated into compressible (where the Fermi energy is pinned to one of the Landau levels) and incompressible strips (where Fermi energy falls into Landau gap).

The spatial distribution of the current density and the local electron temperature at quantizing magnetic fields at low temperatures in 2DES, are extensively investigated in recent years \cite{Akera2002:228,Ise2005:259,Komiyama2006:045333}. G\"uven and Gerhardts calculated the spatial distribution of the current density $\mathbf{j}_{n_{\mathrm{el}}}(x,y)$ and position-dependent electrochemical potential $\mu_{\mathrm{ec}}(x,y)$ utilizing a local version of the Ohm's law, starting from the electrostatic quantities obtained within a self-consistent Thomas-Fermi-Poisson approximation (TFPA) to include classical electron-electron (Hartree) interaction \cite{Guven2003:115327}. This approach is known as the screening theory of the integer quantized Hall effect and will be briefly discussed above. In screening theory, heating effects were neglected hence the electron temperature $T_{\mathrm{e}}$ is assumed to be uniform through the system and equals to the lattice temperature $T_{\mathrm{L}}$. Subsequently Akera and his co-workers implemented the above mentioned calculation scheme, however, also by solving the thermo-hydrodynamical equations they were able to describe dissipative effects and thereby can obtain the spatial distribution of the electron temperature. They showed that, the current carrying incompressible strips strongly affect the electron temperature. The electrostatic boundary conditions imposed in this work are viable mainly for an homogeneous system, however, becomes questionable when considering realistic samples. In any case, this approach opens a new window to investigate the heating effects at a microscopic level and provides reasonable estimates when compared to experiments.

In this paper, we investigate the spatial dependence of the electron temperature in quantum Hall systems (QHS), where compressible (CS) and incompressible strips (IS) are present as a result of interactions \cite{Siddiki2004:195335,Siddiki2003:125315,Siddiki2007:045325}. We study the spatial distribution of the local electron temperature employing a theory of thermo-hydrodynamics in linear response regime, in which the current-induced change of the electrostatic potential equals that of the electrochemical potential, so the induced current does  not influence the electron distribution \cite{Siddiki2004:3541}. The theory of thermo-hydrodynamics is bound by conservation of electron number and thermal flux densities, following Akera and his co-workers \cite{Akera2000:3174,Akera2001:1468}.
Our investigation differs from these pioneering works, as we impose realistic boundary conditions and therefore describe experimental systems accurately. This enables us to predict an unexpected behavior that can be observed in the QHS based Aharonov-Bohm interference experiments, which is dictated by the local electron temperature variations.

\section{Model}\label{model}
\subsection{Basics of the screening theory}

We consider a 2DES in the $z=0$ plane, which is subjected to a perpendicular magnetic field $\mathbf{B}=(0,0,B)$, together with a translational invariance in $x$ direction.

We assume that the electrons are confined by the background potential $V_{\mathrm{bg}}(y)$ generated due to ionized donors, which are distributed uniformly in the $xy$ plane. The local electron (number) density is described by $n_{\mathrm{el}}(y)$. To describe the experimental geometries, we impose boundary conditions such that two metallic gates reside at the physical edges, following Chklovskii et al. The effective potential within the semi-classical approximation can be written as
\begin{equation}\label{1a}
V(y)=V_{\mathrm{bg}}(y)+V_{\mathrm{H}}(y)
\end{equation}
with the confinement potential
\begin{equation}\label{2a}
V_{\mathrm{bg}}(y)=-E_{\mathrm{bg}}^{0}\sqrt{1-\left(\frac{y}{d}\right)^{2}}, \quad \quad E_{\mathrm{bg}}^{0}=\frac{2\pi e^{2}}{\kappa}n_{0}d
\end{equation}
where $E_{\mathrm{bg}}^{0}$ is the pinch-off energy which defines the minimum of the bare confinement potential \cite{Siddiki2006:34}. Hartree potential follows as
\begin{equation}\label{3a}
V_{\mathrm{H}}(y)=\frac{2e^{2}}{\kappa}\int^{d}_{-d}dy^{'}K(y,y^{'})n_{\mathrm{el}}(y^{'}).
\end{equation}
Here $\kappa$ is the dielectric constant (12.4 for GaAs), $n_{0}$ is the donor density and $2d$ is the sample width. Kernel $K(y,y^{'})$ solves the Poisson's equation considering the above mentioned boundary conditions
\begin{equation}\label{4a}
K(y,y^{'})=\ln\left|\frac{\sqrt{(d^{2}-y^{2})(d^{2}-y^{'2})}+d^{2}-y^{'}y}{(y-y^{'})d}\right|.
\end{equation}
Note the fact that the boundary conditions we imposed results in different kernels compared to Ref. \cite{Kanamaru2006:064701,Akera2001:1468} and affect the strength of interactions considerably.

The electron density is, in turn, determined by the effective potential $V(y)$ and is calculated within the Thomas-Fermi approximation
\begin{equation}\label{5a}
n_{\mathrm{el}}(y)=\int dE D(E) f(E+V(y)-\mu_{\mathrm{ec}}),
\end{equation}
where $D(E)$ is the density of states (DOS), $f(E)$ is the Fermi function and $\mu_{\mathrm{ec}}$ is the electrochemical potential \cite{Guven2003:115327,Siddiki2004:195335,Gerhardts2008:245}.

In this work, we apply the local equilibrium approximation, used commonly to describe similar systems \cite{Akera2005:997}. In local equilibrium, the energy distribution of an electron is defined by the Fermi function
\begin{equation}\label{3}
f(\varepsilon,\mu_{\mathrm{ec}},T_{\mathrm{e}})=\frac{1}{\{\mathrm{exp}[(\varepsilon-\mu_{\mathrm{ec}})/k_{\mathrm{B}}T_{\mathrm{e}}]+1\}},
\end{equation}
where $\varepsilon$ is the energy and $T_{\mathrm{e}}$ is the electron temperature. In local equilibrium approximation, the lattice temperature $T_{\mathrm{L}}$ remains unchanged in the presence of an applied current.
If an external current is imposed the electrochemical potential $\mu_{\mathrm{ec}}(\mathbf{r})$ depends on position and its gradient $\mathbf{\mathbf{E}}=\mathbf{\nabla} \mu_{\mathrm{ec}}(\mathbf{r})/e$ satisfies the local Ohm's law
 \begin{equation}\label{7a}
\hat{\rho}(\mathbf{r})\mathbf{j}_{n_{\mathrm{el}}}(\mathbf{r})=\mathbf{E}(\mathbf{r}),
 \end{equation}
hence local current densities can be obtained if local resistivities are provided. We utilize the self-consistent Born approximation (SCBA) to describe both the local conductivities and density of states in the presence of an external magnetic field, similar to Ref.~\cite{Siddiki2004:195335}. Note that, the dissipative current $I$ is the integral of current densities over the sample,
 \begin{equation}\label{6a}
 I=\int_{-d}^{d}dyj_{n_{\mathrm{el}}x}(x,y).
 \end{equation}

\begin{figure}[h!]
\begin{center}\leavevmode
\includegraphics[width=9cm,height=7cm]{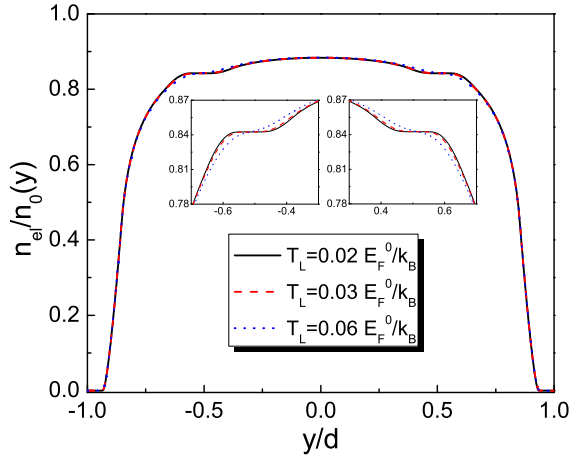}
\caption{Density profile for different lattice temperatures and the magnetic field $\hbar\omega_{c}/E_{F}^{0}=\Omega_{c}/E_{F}^{0}=0.95$
($E_{F}^{0}= 12.68$ meV) with the sample width of $d=1.6$ $\mu$m and the donor density $n_{0}=3.61\times 10^{11}\mathrm{cm}^{-2}$. The insets show the enlarged plateau regions (incompressible strips).}
\label{fig:figure1}
\end{center}
\end{figure}
While presenting our results we use the Fermi energy at the center $E_{F}^{0}$ of the 2DES as the energy scale, since this quantity remains constant once the sample properties (e.g. depletion length, $n_0$ etc.) are fixed, whereas the cyclotron frequency $\omega_{c}=eB/(mc)$ is used to denote the field strength. The magnetic length $\ell=\sqrt{eB/\hbar}$ provides a length scale, quantifying the importance of quantum mechanical effects such as the width of the wave function. As commonly used, the filling factor $\nu=2\pi\ell^2n_{\mathrm{el}}$ describes the occupation of the Landau levels. If electron density exactly equals to the magnetic flux (number) density the Landau level is fully occupied and $\nu$ is an integer. This situation is called an incompressible state, where no states are available at the Fermi energy. Hence, within an incompressible state, electron density distribution is constant. It is straight forward to define the local version of the filling factor by $\nu(x,y)=2\pi\ell^2n_{\mathrm{el}}(x,y)$.

A typical result of the above described calculation scheme is shown in Fig.\ref{fig:figure1}, where the electron density distribution is plotted considering three different lattice temperatures. We fixed the $B$ field strength as $\hbar\omega_{c}/E_{F}^{0}=\Omega_{c}/E_{F}^{0}=0.95$, where the bulk filling factor $\nu(0)$ is slightly above 2. Here, Fermi energy at the bulk is calculated to be 12.68 meV. At the highest lattice temperature ($k_{\mathrm{B}}T_{\mathrm{L}}/E_{F}^{0}=0.06$), the electronic system is in a pure compressible state since both the lowest and next Landau levels are partially occupied due to the fact that the thermal energy of the electrons exceed the Landau gap. Therefore, no incompressible strips are formed. Once lowering the temperature, we observe that density distribution presents plateaus (i.e. is constant within a finite spatial interval) at opposing edges, which are the expected incompressible strips, where Fermi energy falls within the Landau gap locally hence, screening is poor and electrons cannot be redistributed. For the lowest lattice temperature, the incompressible strips become even larger and stable.

\begin{figure}[h]
\begin{center}\leavevmode
 \includegraphics[width=9cm,height=7cm]{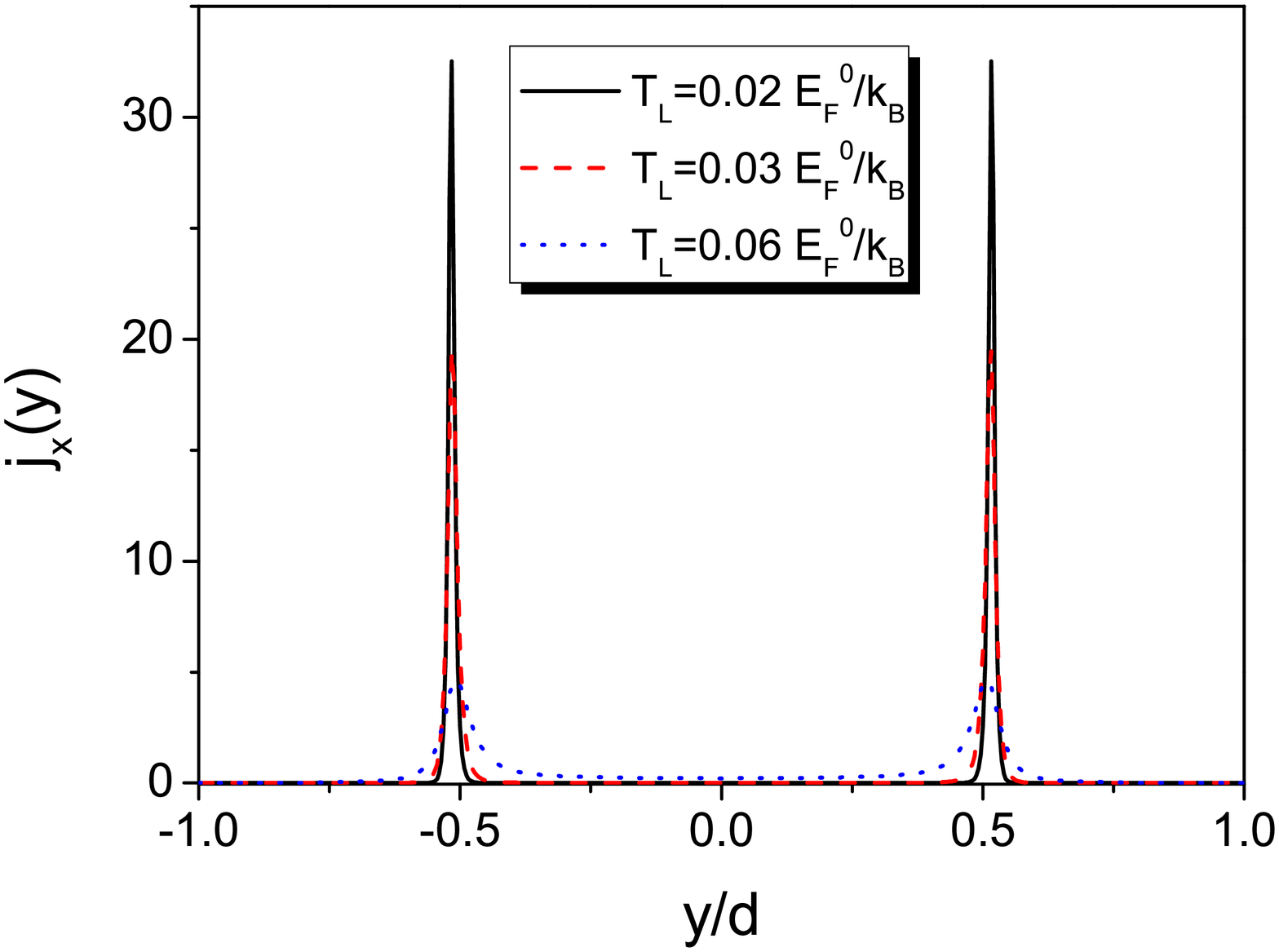}
 \caption{The current densities $j_{x}$ versus position, for different values of the lattice temperature, $T_{\mathrm{L}}=0.02, 0.03$ and $0.06$ $E_{F}^{0}/k_{\mathrm{B}}$ at fixed magnetic field, $\Omega_{c}/E_{F}^{0}=0.95$. Sample parameters are $d=1.6$ $\mu$m and $n_{0}=3.61\times 10^{11}$cm$^{-2}$.}
 \label{fig:figure2}
 \end{center}
 \end{figure}

Once the local electron distribution is known, using the Ohm's law and SCBA to obtain local conductivities one can calculate the current densities within the linear response regime where the imposed current does not effect the electron distribution itself. In Fig.~\ref{fig:figure2}, we show the calculated current density distributions considering the filling factor distribution in Fig.~\ref{fig:figure1}. One can clearly see that, at the lowest lattice temperature all the imposed current is confined to the incompressible strips, where scattering is completely suppressed due to the lack of available states. However, while increasing the temperature the current starts to flow also from the bulk. From these results, one can conclude that the effect of temperature is decisive both determining the electron and current densities. Therefore, in the next step we investigate the effect of imposed current on the electron temperature via solving the thermo-hydrodynamical equations.
\subsection{Thermo-hydrodynamical equations}

We consider two hydrodynamic equations and assume that the electron number and the total energy of the system at hand is conserved. The conservation of the electron number is given by
\begin{equation}\label{4}
\frac{\partial n_{\mathrm{el}}}{\partial t}=-\nabla . \mathbf{j}_{n_{\mathrm{el}}},
\end{equation}
where the number flux density is $\mathbf{j}_{n_{\mathrm{el}}}$. The energy conservation is formulated by
\begin{equation}\label{5}
\frac{\partial \varepsilon}{\partial t}=-\nabla . \mathbf{j}_{\varepsilon}-P_{\mathrm{L}},
\end{equation}
with the energy flux density  $\mathbf{j}_{\varepsilon}$ and the energy loss per unit area $P_{\mathrm{L}}$  due to the heat transfer between electrons and phonons \cite{Akera2005:997}. The time evolution of the entropy density $s$ is derived by using Eqs. (\ref{4}), (\ref{5}), and by the fundamental thermodynamical equation
\begin{equation}\label{6}
T_{\mathrm{e}}ds=d\epsilon-\mu_{\mathrm{\mathrm{ec}}}dn_{\mathrm{el}},
\end{equation}
that yields
\begin{equation}\label{7}
T_{\mathrm{e}}\frac{\partial s}{\partial t}=-\nabla.\mathbf{j}_{q}-\nabla\mu_{\mathrm{ec}}.\mathbf{j}_{n_{\mathrm{el}}}-P_{\mathrm{L}},
\end{equation}
where the thermal flux density $\mathbf{j}_{q}$ is described by
\begin{equation}\label{8}
 \mathbf{j}_{q}= \mathbf{j}_{\varepsilon}-\mu_{\mathrm{ec}}\mathbf{j}_{n_{\mathrm{el}}}.
\end{equation}
Utilizing the above equations that describe the drift and hopping processes, the number flux densities $\mathbf{j}_{n_{\mathrm{el}}}$ and the thermal flux densities $\mathbf{j}_{q}$ can be written as \cite{Akera2005:997},
\begin{equation}\label{9}
 j_{n_{\mathrm{el}}}(x)= -L_{xx}^{11}\nabla_{x}\mu_{\mathrm{ec}}+L_{yx}^{11}\nabla_{y}V-L_{xx}^{12}T_{\mathrm{e}}^{-1}\nabla_{x}T_{\mathrm{e}},
\end{equation}
\begin{equation}\label{10}
 j_{n_{\mathrm{el}}}(y)= -L_{yx}^{11}\nabla_{x}V-L_{xx}^{11}\nabla_{y}\mu_{\mathrm{ec}}-L_{xx}^{12}T_{\mathrm{e}}^{-1}\nabla_{y}T_{\mathrm{e}},
\end{equation}
\begin{equation}\label{11}
 j_{q}(x)= -L_{xx}^{12}\nabla_{x}\mu_{\mathrm{ec}}+K_{yx}^{21}\nabla_{y}V-L_{xx}^{22}T_{\mathrm{e}}^{-1}\nabla_{x}T_{\mathrm{e}},
\end{equation}
\begin{equation}\label{12}
 j_{q}(y)= -K_{yx}^{21}\nabla_{x}V-L_{xx}^{12}\nabla_{y}\mu_{\mathrm{ec}}-L_{xx}^{22}T_{\mathrm{e}}^{-1}\nabla_{y}T_{\mathrm{e}}.
\end{equation}
Here, we use the transport coefficients (i.e. $L_{jk}$ and $K_{jk}$, where $\{j,k\}$ denote $x$ and $y$) given by Kanamaru \emph{et al} \cite{Kanamaru2006:064701}.

As we assume translation invariance in the $x$ direction, the electron temperature $T_{\mathrm{e}}$ and the electrochemical potential $\mu_{\mathrm{ec}}$ do not depend on $x$.  Taking into account these boundary conditions, conservation equations reduce to
\begin{equation}\label{16}
\Delta j_{n_{\mathrm{el}}}(y)=0 \quad \quad (|y|<d),
\end{equation}
\begin{equation}\label{17}
\nabla_{y}(\Delta j_{q}(y))+eE_{x}j_{n_{\mathrm{el}}}(x)+P_{\mathrm{L}} =0,
\end{equation}
where
\begin{equation}\label{18}
\Delta \mathbf{j}_{n_{\mathrm{el}}}=\mathbf{j}_{n_{\mathrm{el}}}-\mathbf{j}_{n_{\mathrm{el}}}^{\mathrm{eq}},
\end{equation}
\begin{equation}\label{19}
\Delta \mathbf{j}_{q}=\mathbf{j}_{q}-\mathbf{j}_{q}^{\mathrm{eq}}.
\end{equation}
Here $\mathbf{j}_{n_{\mathrm{el}}}^{\mathrm{eq}}$ and $\mathbf{j}_{q}^{\mathrm{eq}}$ are the number flux density and the thermal flux density in equilibrium \cite{Kanamaru2006:064701}.

One can see that, if the spatial distributions of the electrostatic and electrochemical potentials are known the above equations can be solved to obtain the local electron temperatures. In order to perform calculations for a finite width sample, we consider a hard wall potential ($V(y)=\infty$ for $|y|>d$). Therefore flux is absent outside of the sample. We integrate Eqs. (18) and (19) within a narrow region on the boundary to
obtain the boundary conditions. These conditions do not influence the distribution of electrons within the sample.

In the following we present results, where the electron temperature is obtained starting from the results of the screening theory. Note that, the imposed current is assumed to be sufficiently small to guarantee that system is still within the linear response regime, hence, both the induced potential and the variations of the local electron temperature do not influence the electronic distribution.

 \begin{figure}[h]
\begin{center}\leavevmode
 \includegraphics[width=9cm,height=7cm]{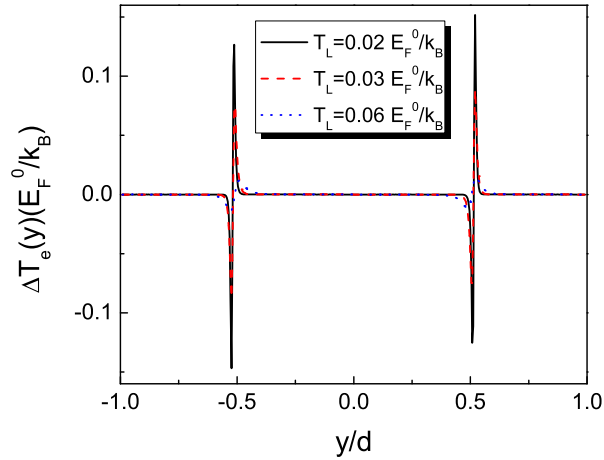}
 \caption{Calculated the local electron temperature deviation $\Delta T_{\mathrm{e}}$ versus position, for different values of the lattice temperature, $T_{\mathrm{L}}=0.02, 0.03$ and 0.06 $E_{F}^{0}/k_{\mathrm{B}}$. Sample parameters are $d=1.6$ $\mu$m, $n_{0}=3.61\times 10^{11}$cm$^{-2}$ and $\Omega_{c}/E_{F}^{0}=0.95$.}
 \label{fig:figure3}
 \end{center}
 \end{figure}

The local electron temperature variation as a function of spatial coordinate is shown in Fig.\ref{fig:figure3}, where the electron and current density profiles are shown in Figs.~\ref{fig:figure1} and \ref{fig:figure2}, respectively. We observe that, assuming a higher lattice temperature results in small variations at the local electron temperature. This is expected, since at higher temperatures the incompressible strips are not well developed, hence, current is spread all over the sample. Therefore, heating effects take place almost at each location. Most interestingly, one observes that one side of the sample heats up, whereas the opposing edge is cooled down. This effect is nothing but the Peltier effect in which cooling occurs at the junction of two dissimilar thermoelectric materials when an electric flows through the junction. Peltier heat is proportional to the current and changes sign if the current direction is reversed \cite{Ahlswede01:562,Ahlswede02:165}. Although there is no net charge transfer in the $y$ direction the thermal flux is transferred by the conservation of energy. Lowering the lattice temperature yields stronger incompressible strips, hence, current is more confined to these regions. As a direct consequence, the local electron temperature starts to vary stronger.

\begin{figure}[h!]
\begin{center}\leavevmode
 \includegraphics[width=9cm,height=7cm]{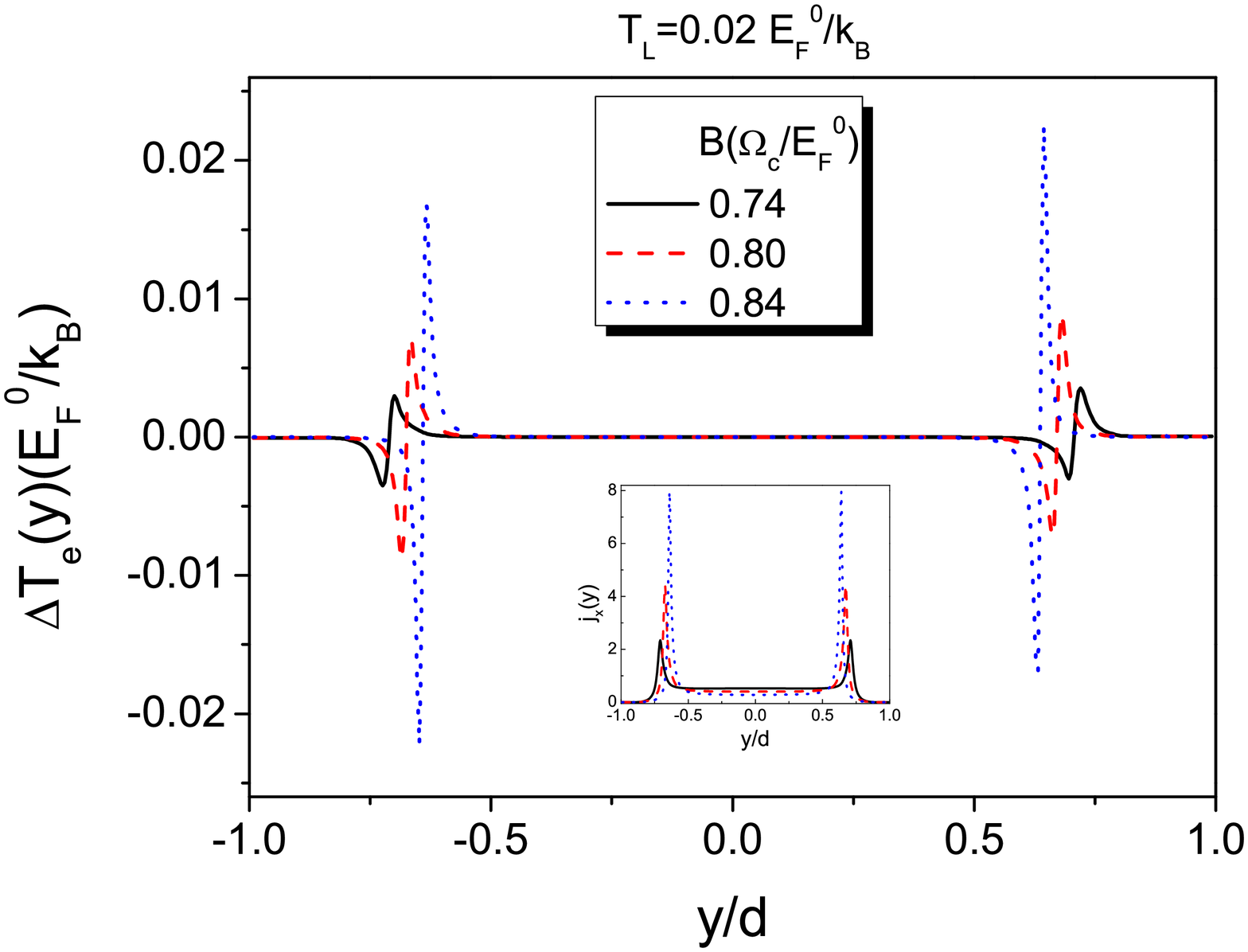}
 \caption{Calculated the local electron temperature deviation $\Delta T_{\mathrm{e}}$ versus position, for different values of the magnetic field, $\Omega_{c}/E_{F}^{0}=0.74, 0.8$ and $0.84$. Sample parameters are $d=1.6$ $\mu$m, $n_{0}=3.61\times 10^{11}$cm$^{-2}$ and $T_{\mathrm{L}}=0.02$ $E_{F}^{0}/k_{\mathrm{B}}$. The inset shows the current density profile.}
 \label{fig:figure4}
 \end{center}
 \end{figure}

Fig.~\ref{fig:figure4} shows the local electron temperature deviation $\Delta T_{\mathrm{e}}$ as a function of the magnetic field with constant lattice temperature $T_{\mathrm{L}}$. It is seen that  $\Delta T_{\mathrm{e}}$ shows spatially antisymmetric behavior that is obviously due to thermal flux transfer in the $y$ direction. Note also that, the variation of the local electron temperature strongly depends on magnetic field. Since, the widths of the incompressible strips decreases by decreasing the magnetic field and the strips move closer to the sample edge.

\section{Conclusion}
In this paper, we calculated the local electron temperature and the current density profiles considering a quantum Hall system. First the position dependent electrostatic and electrochemical potentials are calculated within the Thomas-Fermi-Poisson approximation, self-consistently. Next, we used these quantities to calculate the local electron temperature in the linear response regime, utilizing the energy and particle conservation when solving the thermo-hydrodynamical equations. We found that, the current carried by the incompressible strips heats the electron system locally. Hence at low temperatures and at the low field side of the quantized Hall plateau the heating is mainly local, whereas at the high field end of the plateau heating effect is spread all over the sample. This situation is altered at high lattice temperatures, since there are no well developed incompressible strips. Interestingly, we observed that one side of the sample heats up, whereas the opposite side is cooled down. From these observations, we expect that the edge-state transport is highly sensitive to formation of the incompressible strips. This leads us to conclude that, even at low currents local heating effects may become important. As an interesting example, we claim that the uncovered dephasing processes at the Aharonov-Bohm type interference experiments would yield a quantitative test for heating effects. We expect that, at low temperatures and at the low-field side of the plateaus the visibility of the interference signal is strongly influenced by the local heating effects. A simple test would be to compare the fading of the visibility as a function of applied current, measured at sharp edge defined (e.g. by trench gating) samples and soft edge defined (e.g. gate) samples. Since, the incompressible strips are more stable at gate defined samples, we predict that, heating effects will be more effective at sharp edge samples. Therefore, the visibility should fade away at the sharp edge samples faster compared to gate defined samples.

\section*{Acknowledgments}
 This work was partially supported by the Scientific and Technical Research Council of Turkey (TUBITAK) under Grant no. 109T083, IU-BAP:6970.
\bibliographystyle{elsarticle-num}

\begin{thebibliography}{10}
%

\bibitem{Klitzing80:449}  K. von Klitzing, G. Dorda, and M. Pepper, New method for high-accuracy determination of the fine-structure constant based on quantized Hall, Phys. Rev. Lett. 45 (1980) 494-497.

\bibitem{FQHE} D.C. Tsui and H.L. Stormer and A.C. Gossard, Two-dimensional magnetotransport in the extreme quantum limit, Phys. Rev. Lett. 48 (1982)
1559–1562.

\bibitem{Ebert83:5441}  G. Ebert, K. von Klitzing, K. Ploog, and G. Weimann, Two-dimensional magneto-quantum transport on GaAs$-$Al$_{x}$Ga$_{1-x}$As heterostructures under non-ohmic conditions,  J. Phys. C: Solid State Phys.  16 (1983) 5441.

\bibitem{Cage83:1374}  M. E. Cage, R. F. Dziuba, B. F. Field, E. R. Williams, S. M. Girvin, A. C. Gossard,
 D. C. Tsui, and R. J. Wagner, Dissipation and dynamic nonlinear behavior in the quantum Hall regime,  Phys. Rev. Lett. 51 (1983) 1374-1377.

\bibitem{Kuchar84:196}  F. Kuchar, G. Bauer, G. Weimann, and H. Burkhard, Non equilibrium behavior of the two-dimensional electron gas in the quantized Hall resistance regime of GaAs/Al$_{0.3}$Ga$_{0.7}$As, Surf. Sci. 142 (1984) 196.

\bibitem{Kaya98:7536}  I. I. Kaya, G. Nachtwei, K. von Klitzing, and K. Eberl, Spatial evolution of hot-electron relaxation in quantum Hall conductors, Phys. Rev. B 58 (1998) 7536-7539.

\bibitem{Akera2002:228}  H. Akera, Hydrodynamic equations in quantum Hall systems at large currents, J. Phys. Soc. Jpn. 71 (2002) 228-236.

\bibitem{Weitz00:247} P. Weitz, E. Ahlswede, J. Weis, K. von Klitzing, and K. Eberl, Hall-potential investigations under quantum Hall conditions using scanning force microscopy, Physica E 6  (2000) 247-250.



\bibitem{Dahlem:2010:121305} F. Dahlem, E. Ahlswede,y J. Weis, and K. v. Klitzing, Cryogenic scanning force microscopy of quantum Hall samples: Adiabatic transport originating in anisotropic depletion at contact interfaces, Phys. Rev. B 82 (2010)  121305.


\bibitem{Ise2005:259}  T. Ise, H. Akera and H. Suzuura, Electron temperature distribution and hot spots in quantum Hall systems, J. Phys. Soc. Jpn. 74 (2005) 259-262.

\bibitem{Komiyama2006:045333}  S. Komiyama, H. Sakuma, and K. Ikushima, Electron temperature of hot spots in quantum Hall devices, Phys. Rev. B 73 (2006) 045333.   

\bibitem{Guven2003:115327} K. G\" uven and R. R. Gerhardts, Self-consistent local equilibrium model for density profile and distribution of dissipative currents in a Hall bar under strong magnetic fields, Phys. Rev. B 67 (2003) 115327.

\bibitem{Siddiki2004:195335}  A. Siddiki and R. R. Gerhardts, Incompressible strips in dissipative Hall bars as origin of quantized Hall plateaus, Phys. Rev. B 70 (2004) 195335.



\bibitem{Siddiki2003:125315}  A. Siddiki and R. R. Gerhardts, Thomas-Fermi-Poisson theory of screening for laterally confined and unconfined two-dimensional electron systems in strong magnetic fields, Phys. Rev. B 68 (2003) 125315.   

\bibitem{Siddiki2007:045325}  A. Siddiki and F. Marquardt, Self-consistent calculation of the electron distribution near a quantum point contact in the integer quantum Hall effect, Phys. Rev. B 75 (2007) 045325.  

\bibitem{Siddiki2004:3541}  A. Siddiki and R. R. Gerhardts, The interrelation between incompressible strips and quantized Hall plateaus, International Journal of Modern Physics B 18 (2004) 3541.  


\bibitem{Akera2000:3174}  H. Akera, Electronic processes at the breakdown of the quantum Hall effect, J. Phys. Soc. Jpn. 69  (2000) 3174-3177.

\bibitem{Akera2001:1468}  H . Akera, Hydrodynamic equation for the breakdown of the quantum Hall effect in a uniform current, J. Phys. Soc. Jpn. 70 (2001)  1468-1471.

    \bibitem{Siddiki2006:34}  A. Siddiki, S. Kraus and R. R. Gerhardts, Screening model of magneto-transport hysteresis observed in bilayer quantum Hall systems, Physica E 34 (2006) 136-139.   

\bibitem{Kanamaru2006:064701}    S.Kanamaru, H. Suzuura and H. Akera, Spatial Distributions of Electron Temperature in Quantum Hall Systems with Compressible and Incompressible Strips, J. Phys. Soc. Jpn. 75 (2006) 064701.




\bibitem{Gerhardts2008:245}  R. R. Gerhardts, The Effect of Screening on Current Distribution and Conductance Quantisa- tion in Narrow Quantum Hall Systems,  Phys. Stat. Sol. (b) 245 (2008) 378-392.

\bibitem{Akera2005:997}  H. Akera and  H. Suzuura, Thermohydrodynamics in quantum Hall systems, J. Phys. Soc. Jpn. 74 (2005) 997-1005.

\bibitem{Ahlswede01:562} E. Ahlswede and P. Weitz and J. Weis and K. von Klitzing and K. Eberl, Hall potential profiles in the quantum Hall regime measured by a scanning force microscope, Physica B 298 (2001) 562-566.

\bibitem{Ahlswede02:165} E. Ahlswede and J. Weis and K. von Klitzing and K. Eberl, Hall potential distribution in the quantum Hall regime in the vicinity of a potential probe contact, Physica E 12 (2002) 165-168.






\end{thebibliography}

\end{document}